\documentclass{elsart}
\newcommand \bea{\begin{eqnarray}}
\newcommand \eea{\end{eqnarray}}

\newcommand{\av}[1]{\langle{#1}\rangle}

\newcommand{\be}{\begin{eqnarray}}
\newcommand{\ee}{\end{eqnarray}}

\newcommand{\ave}[1]{\left\langle #1 \right\rangle}

\def\bea{\begin{eqnarray}}
\def\eea{\end{eqnarray}}
\def\be{\begin{eqnarray}}
\def\ee{\end{eqnarray}}
\def\ave#1{{\langle #1 \rangle}}

\def\beq{\begin{equation}}
\def\eeq{\end{equation}}

\def\bea{\begin{eqnarray}}
\def\eea{\end{eqnarray}}
\def\ba{\begin{array}}                  
\def\ea{\end{array}}

\begin{document}

\begin{center}

{\bf Net Charge and Isospin Fluctuations in the World of
Elementary Particles }\\
Vesna Mikuta-Martinis\\
Rudjer Boskovic Institute\\
10001 Zagreb, Bijenicka c. 54, P.O.Box 1016, Croatia\\
vmikuta@rudjer.irb.hr

\end{center}

\vspace{2cm}

Charge and isospin  event-by-event fluctuations in high-energy
pp-collisions are predicted within the Unitary Eikonal Model ({\bf
UEM }), in particular the fluctuation patterns of the ratios of
charged-to-charged and neutral-to-charged pions. These
fluctuations are found to be sensitive to the presence of unstable
resonances, such as  $\rho $ and $\omega $ mesons.We predict that
the charge-fluctuation observable $D_{UEM}$ should be restricted
to the interval $8/3\le D_{UEM}\le 4$ depending on the $\rho /\pi
$ production ratio.This result is shown to be compatible with the
formation of a hadron gas at RHIC and SPS energies.

\newpage

\section{Introduction}

A large number of particles produced in a single central
ultrarelativistic heavy-ion collision at RHIC (Relativistic Heavy
Ion Collider)and LHC (Large Hadron Collider) gives us a remarkable
opportunity to study fluctuations of specific hadronic observables
on the event-by-event basis [Blume,2002]. High energy
hadron-nucleus, nucleus-nucleus and heavy-ion collisions may be
treated as a linear superposition of pp-collisions [Bialas {\sl et
al.},1976].

\section{Charge Fluctuations}

The event-by-event fluctuations of the ratio of positively and
negatively charged pions can be a distinct signal of Quark Gluon
Plasma (QGP) formation [Jeon  \& Koch,2000] (in the QGP phase the
unit of charge is 1/3 while in the hadronic phase the unit of
charge is 1).

The fluctuations of the total charge multiplicity $ N_{ch} = N_{+}
+ N_{-} $, and the net charge  $ Q = N_{+}- N_{-} $, are defined
[Heiselberg \& Jackson, 2001] as

\bea
 && \frac{\av{(N_+\pm N_-)^2}-\av{N_+\pm N_-}^2}{\av{N_++ N_-}}
 = \nonumber\\
 &&\quad \frac{\av{N_+}}{\av{N_{ch}}}\omega_{N_+}
    +\frac{\av{N_-}}{\av{N_{ch}}}\omega_{N_-} \, \pm \, C \,, \label{oNpm}
\eea

where the correlation is given by

\bea
  C = \frac{\av{N_+N_-}-\av{N_+}\av{N_-}}{\av{N_{ch}}/2} \,. \label{C}
\eea

In experiments, $\omega_{N_+}\approx \omega_{N_-}$, so that the
fluctuation in total charge simplifies to \bea
  \omega_{N_{ch}} &\equiv& \frac{\av{N_{ch}^2}-\av{N_{ch}}^2}{\av{N_{ch}}}
   =\omega_{N_+} + C \,, \label{oNch}
\eea and that for the net charge becomes \bea
  \omega_Q &\equiv& \frac{\av{Q^2}-\av{Q}^2}{\av{N_{ch}}}
   =\omega_{N_+} - C \,. \label{oQ}
\eea

The ratio of particle multiplicities is considered to avoid the
volume fluctuations  $ (\pi^{+}/\pi^{-}, K /\pi)$. A suitable
observables define by these ratios are $ F = Q / N_{ch} $  and $R=
N_{+}/ N_{-}$.

 If $\ave{N_{\rm ch}} \gg \ave{Q}$ then
 \be
 \ave{\delta R^2}
 =
 \ave{R^2} - \ave{R}^2
 \approx
 4\ave{\delta F^2}
 \ee
 and
 \be
 \ave{\delta F^2}
 =
 {\ave{Q}^2\over \ave{N_{\rm ch}}^2}
 \ave{
 \left(
 {\delta Q\over {\ave Q}}
 -
 {\delta N_{\rm ch}\over {\ave N_{\rm ch}}}
 \right)^2
 }
 \ee
 The fluctuations of the ratio $ R = N_{+} / N_{-} $ are connected to the fluctuations
 in the net charge through the observable
 \be
 D \equiv \ave{N_{\rm ch}}\ave{\delta R^2} = 4\ave{N_{\rm ch}}\ave{\delta F^2}
 =4 {\ave{\delta Q^2} \over \ave{N_{\rm ch}}}
 \label{eq:Dpm}
 \ee
which provides a measure of the charge fluctuations per unit
entropy.

Because of the effects of the finite net charge and the finite
acceptance window in the experiments the observable D should be
corrected (UrQMD - Ultra-relativistic Quantum Molecular Dynamics
model [Bleicher {\sl et al.}, 2000])

\be D \Longrightarrow \tilde{D}={\ave{N_{\rm ch}}_{\Delta y}
\ave{\delta R^2}_{\Delta y}\over C_\mu\, C_y} \ee

and

\bea C_{\mu} = \tilde{R}^2_{\Delta y}= {\ave{N_+}^2_{\Delta y}
\over \ave{N_-}^2_{\Delta y}} \label{eq1} \eea

\be C_y = 1 - P = 1-\frac{\langle N_{\rm ch}\rangle_{\Delta y}}
{\langle N_{\rm ch}\rangle_{\rm total}}\quad. \label{eq2} \ee

 We have calculated observable D within the Unitary Eikonal Model
(UEM)
[Martinis {\sl et al.} 1994], [Martinis {\sl et al.} 1995]
and compared it with the predictions of the termal model, UrQMD
model for hadron gas, and Quark-Gluon gas model.

\section{Unitary Eikonal Model}

The UEM is based on the fact that at high energies most of the
pions are produced in the nearly baryon-free central region of the
phase space. The energy available for their production is
\begin{equation}
E_{had} = \frac{1}{2} \sqrt{s} - E_{leading}
\end{equation}
which at fixed total c.m. energy $ \sqrt{s} $ varies from
event-to-event.

The coherent production of  pions or clusters of pions in the
impact parameter space of leading particles is described by the
factorized form of the scattering amplitude. It is  characterized
by the product of classical source functions describing the
production of a cluster of pions with a definite isospin index.
The cluster decays into pions outside the region of strong
interactions (i.e. the final-state interaction between pions is
neglected.

Studies of nucleus-nucleus and heavy-ion collisions at the
partonic level suggest that the central region is mainly dominated
by gluon jets. Since gluon's isospin is zero, it is very likely
that total isospin of the produced pion cloud in the central
region is also zero. We assume global conservation of isospin.

We consider the cloud of $N$ pions produced in pp-collisions which
consists of $N_{\pi}$ directly produced pions and $2N_{\rho }$
pions  produced via $\rho $-type clusters such that $N = N_{\pi} +
2N_{\rho }$. The probability distribution of producing $N_{+}, \,
N_{ \_} $ and $N_{0}$ pions in the impact parameter space, such
that $N = N_{+} + N_{ \_} + N_{0}$, is

\begin{eqnarray}
P_{I I_{3}}(N_{+}N_{ \_}N_{0}\mid N)  C_{I I_{3}}(N) &  =  & \nonumber \\
& & \hspace*{-4cm} \sum_{I'I'_{3}}\omega_{I', I_{3}'} \int d^{2}b
dq_{1} dq_{2} \ldots dq_{N} \mid \langle I'I'_{3},N_{+}N_{
\_}N_{0} \mid \hat{S}(s, \vec{b}) \mid II_{3} \rangle \mid^{2})
\end{eqnarray}
where
\begin{eqnarray}
 N &  =  & N_{+} + N_{ \_} + N_{0}\nonumber \\
 N_{+} &  = & n_{\pi ^+} + n_{\rho ^+} + n_{\rho ^0} \nonumber\\
 N_{-} &  = & n_{\pi ^-} + n_{\rho ^-} + n_{\rho ^0} \nonumber \\
 N_{0} &  = & n_{\pi ^0} + n_{\rho ^+} + n_{\rho ^-}
\end{eqnarray}

\section{Results}

We predict that the charge-fluctuation observable  $ D_{UEM}$
should be restricted to the interval

\begin{equation}
 8/3\le D_{UEM}\le 4
\end{equation}
depending on the $\rho /\pi $ production ratio. The upper bound
$D_{UEM} = 4$ is satisfied if $\overline{n}_{ \rho} = 0$. In that
case the pion production is restricted only by the global
conservation of isospin. However, if $\overline{n}_{ \pi} = 0$ the
lower limit is $D_{UEM} = 8/3$.
 The predictions of the UrQMD model are

\be
 \tilde{D}
  =
 {\ave{N_{\rm ch}}_{\Delta y} \ave{\delta R^2}_{\Delta y}\over C_\mu\, C_y}
 =
\left\{
\begin{array}{ll}
1 & \hbox{quark gluon gas}\\
2.8 & \hbox{resonance gas}\\
4 & \hbox{uncorrelated pion gas}
\end{array}
\right. \label{eq:correct}
 \ee
what is in a good agreement with the results of our model
(Eq.(14)).

\section{Conclusion}

Our results are shown to be compatible with the formation of a
hadron gas at RHIC and SPS energies [ Adcox, 2002 ],[ Blume, 2002
],[ Reid, 2002 ] and differ noticeably from that expected for QGP
.We can conclude that there is no indication for
Quark-Gluon-Plasma yet.

\section{Acknowledgements}
This work was supported by the Ministry of Science, Education and
Sports of Croatia under Contract No. 0098004.\\

\section{References}

\hspace{1cm}Adcox, K. for the PHENIX collaboration,[2002] " Net
charge fluctuations in Au + Au Interactions  at $\sqrt{s_{NN}}$ =
130 GeV ", {\sl Phys. Rev. Lett.} {\bf 89},082301. \\

\hspace{1cm}Bialas, A.,Bleszynski, M. \& Czyz, W. [1976]
"Multiplicity distributions in nucleus-nucleus collisions at high
energies ",{\sl Nucl. Phys.} {\bf B111},461-476.\\

\hspace{1cm}Bleicher, M.,Jeon, S.\& Koch, V.[2000] "Event-by-event
fluctuations of the charged particle ratio from non-equilibrium
transport theory",{\sl Phys. Rev.}
{\bf C62},061902.\\

\hspace{1cm}Blume, C. for the NA49 collaboration,[2002] " New
results from NA49", {\sl  Nucl. Phys.} {\bf A698}, 104c-111c.\\

\hspace{1cm}Heiselberg, H. \& Jackson, A. D. [2001] " Anomalous
multiplicity fluctuations from phase transitions in heavy ion
collisions ",{\sl Phys. Rev.} {\bf C63}, 064904.\\

\hspace{1cm}Jeon, S. \& Koch, V. [2000] "Charged particle ratio
fluctuation as a signal for QGP ", {\sl Phys. Rev. Lett.}{\bf 85},
2076-2079.\\

\hspace{1cm}Martinis, M., Mikuta-Martinis, V., Svarc, A. \&
Crnugelj, J.[1994] " Isospin correlations in high-energy heavy-ion
collisions ", {\sl FIZIKA } {\bf B3}, 197-205. \\

\hspace{1cm}Martinis, M., Mikuta-Martinis, V. \& Crnugelj, J.
[1995] " Can pions created in high energy heavy ion collisions
produce a Centauro type effect ?",{\sl Phys. Rev.} {\bf C52},
1073-1075.\\

\hspace{1cm}Reid, J. G. for the STAR collaboration, [2002] "STAR
event-by-event fluctuations", {\sl Nucl. Phys.}{\bf A698},
611c-614c.

\end{document}